\documentclass[lettersize,journal]{IEEEtran}
\usepackage{amsmath,amsfonts}
\usepackage{algorithmic}
\usepackage[linesnumbered,ruled,vlined]{algorithm2e}
\usepackage{array}
\usepackage[caption=false,font=normalsize,labelfont=sf,textfont=sf]{subfig}
\usepackage{textcomp}
\usepackage{stfloats}
\usepackage{url}
\usepackage{verbatim}
\usepackage{graphicx}
\usepackage{cite}
\usepackage{epstopdf}
\usepackage{color, soul}
\SetKwInOut{Input}{Input}
\SetKwInOut{Output}{Output\,}

\begin{document}

\title{\huge A Lattice-Reduction Aided Vector Perturbation Precoder Relying on Quantum Annealing}

\author{Samuel~Winter, Yangyishi~Zhang,~\IEEEmembership{Senior~Member,~IEEE}, Gan~Zheng,~\IEEEmembership{Fellow,~IEEE}, Lajos~Hanzo,~\IEEEmembership{Life~Fellow,~IEEE}
\thanks{S. Winter and Y. Zhang are with Research and Network Strategies, BT Group (Email: \{samuel.2.winter, yangyishi.zhang\}@bt.com), G. Zheng is with University of Warwick (Email: gan.zheng@warwick.ac.uk), and L. Hanzo is with University of Southampton (Email: lh@ecs.soton.ac.uk).}
\thanks{S. Winter and Y. Zhang have been supported by Innovate UK project no. 10031626. L. Hanzo would like to acknowledge the financial support of the EPSRC projects EP/W016605/1, EP/X01228X/1 and EP/Y026721/1 as well as of the European Research Council's Advanced Fellow Grant QuantCom (Grant No. 789028). G. Zheng has been supported by EP/X04047X/1.}
}


\maketitle
\bstctlcite{IEEEexample:BSTcontrol}

\begin{abstract}
Quantum annealing (QA) is proposed for vector perturbation precoding (VPP) in multiple input multiple output (MIMO) communications systems. The mathematical framework of VPP is presented, outlining the problem formulation and the benefits of lattice reduction algorithms. Lattice reduction aided quantum vector perturbation (LRAQVP) is designed by harnessing physical quantum hardware, and the optimization of hardware parameters is discussed. We observe a 5dB gain over lattice reduction zero forcing precoding (LRZFP), which behaves similarly to a quantum annealing algorithm operating without a lattice reduction stage. The proposed algorithm is also shown to approach the performance of a sphere encoder, which exhibits an exponentially escalating complexity.
\end{abstract}

\vspace{-0.5cm}
\section{Introduction}
\IEEEPARstart{M}{ultiple} input multiple output (MIMO) systems form the backbone of wireless communications links; they are capable of improving the links' capacity beyond that of single-antenna communications. Multiple antennas may also be harnessed for supporting multiple users by exploiting the unique user-specific channel impulse responses (CIRs), provided that they are accurately estimated. Then, these coordinated antennas used at the base station (BS) can transmit and receive multiple data streams, whilst compensating the inter-link interferences. More explicitly, when considering the downlink (DL), the users may estimate the DL CIRs and signal it back to the BS. By exploiting the CIR, the BS then applies DL transmit precoding (TPC) to pre-compensate for the DL CIRs about to be encountered.

Nonlinear TPC schemes have been shown to provide better performance than linear schemes, but at the cost of exponentially escalating complexity ~\cite{castanheira_linear_2013,zhang_far-end_2020} versus the number of antennas. In this treatise we conceive a vector perturbation precoder (VPP), which transforms the original TPC design problem of maximizing the signal-to-noise ratio (SNR) into the closest vector problem (CVP). The CVP involves finding a lattice point that is closest to a given target vector that does not lie on the lattice, in an $N_r$ dimensional complex vector space. Further restrictions on the legitimate vector search space are imposed due to the limited linear dynamic range of the radio frequency (RF) chain of each antenna, and the limited symbol-duration available to find a solution at the target symbol rate.

Noisy intermediate scale quantum (NISQ) devices are currently being assessed in terms of their suitability for wireless systems. In this context, a quantum search algorithm was developed for multiuser detection that can match the performance of a classical counterpart ~\cite{ye_quantum_2019}. Although, at the time of writing, implementations using quantum hardware are limited by the available capability of the NISQ devices, they hold the promise of reducing computational complexity. The maximum likelihood (ML) detection problem has been solved for small-scale systems using the quantum approximate optimization algorithm ~\cite{cui_quantum_2022}. Indeed, the employment of quantum annealing (QA) in ML detection is already a well studied problem ~\cite{tabi_evaluation_2021,de_luna_ducoing_quantum_2022}. Tabi \textit{et al.} ~\cite{tabi_evaluation_2021} employ QA in massive MIMO communications, and probe the limits of what is possible based on the available hardware. It is also shown that for a quadrature amplitude modulation (QAM) scheme of order 16, QA was able to match the performance of a classical ML detector. As a further advance, De Luna \textit{et al.} ~\cite{de_luna_ducoing_quantum_2022} demonstrate the ability to parallelize computations for DWave processors, but at the time of writing the number of qubits remains limited. The limitations imposed by the number of bits per symbol are attributed to the resolution of the annealing control. As for DL TPC, Kasi \textit{et al.} present a novel preprocessing technique that relaxes the hardware requirements, for a QA VPP solver, at a modest performance erosion ~\cite{kasi_quantum_2021}. Their experiments show steps taken to mitigate hardware constraints, however a strong dependence on the time allowed for the QA shows a strong susceptibility to noise.

Against this background, a novel QA based framework is proposed for VPP, which substantially improves the robustness against hardware limitations. The algorithm, unlike the previous work, is designed with the physical mechanism of QA in mind. Explicitly, we conceive a powerful preprocessing procedure for the QA by harnessing the Lenstra-Lenstra-Lovasz (LLL) algorithm ~\cite{lenstra_factoring_1982} for reducing the search space of the vectors and adjusting the quantum mechanical form of the problem. Additionally, we incorporate a noise resilient encoder for reducing the impact of quantum hardware impairments on the system performance. The performance of the proposed lattice reduction aided quantum vector perturbation (LRAQVP) scheme is experimentally validated on real quantum hardware as well as benchmarked against its classical lattice reduced zero forcing (LRZFP) counterpart.

Notation: scalar quantities are represented by normal letters, vector quantities by lowercase boldface letters, and matrices by uppercase boldface letters. The Euclidean and Frobenius norms are represented by $\lVert \cdot \rVert$ and $\lVert \cdot \rVert_F$ respectively; the floor operator by $\lfloor \cdot \rfloor$. The (right pseudo) inverse of a matrix is denoted by $\left( \cdot \right)^+$ and the matrix/vector transpose by $\left( \cdot \right)^\top$. The functions $\mathfrak{Re}[\cdot]$ and $\mathfrak{Im}[\cdot]$ extract the real and imaginary parts of its argument.

\vspace{-0.4cm}
\section{System Model}

A BS having $N_t$ transmit antennas serves $N_r$ users, each with a single antenna, where $N_r \le N_t$. The symbol vector $\mathbf{u} \in \mathbb{C}^{N_r}$ contains the symbols transmitted to each user. The elements of $\mathbf{u}$ are drawn from a uniform distribution of complex square QAM symbols; this choice of modulation is however not a requirement for VPP ~\cite{li_constellation_2015}. The average transmit power of a symbol is 1. We define $\rho = \mathcal{E}_s / N_s$ as the average symbol energy $\mathcal{E}_s$ over the noise variance per symbol $N_s$. The received signal vector is
\begin{equation}
        \mathbf{y} = \sqrt{\frac{\rho}{\lVert \mathbf{x} \rVert^2}} \mathbf{H} \mathbf{x} + \mathbf{n},
\end{equation}
where $\mathbf{H} \in \mathbb{C}^{N_r \times N_t}$ represents the channel state information, which is assumed to be perfectly known, and $\mathbf{n} \in \mathbb{C}^{N_r}$ is the noise vector. The elements of $\mathbf{H}$ and $\mathbf{n}$ are drawn from the complex Gaussian distribution $\mathcal{CN}(0, 1)$. Lastly, the precoded transmit vector of
\begin{equation}
        \label{eq_x_def}
        \mathbf{x} = \mathbf{G}(\mathbf{u} - \tau \mathbf{l})
\end{equation}
is dependent on the (pseudo)inverse of $\mathbf{H}^{+} = \mathbf{G}$ and on an integer perturbation vector $\mathbf{l} \in \mathbb{G}^{N_r}$, where $\mathbb{G}$ represents the set of Gaussian integers. The perturbation vector is scaled to match the width of the QAM scheme by $\tau =2 c_{max} +\Delta$, where $\Delta$ represents the distance between two neighbouring constellation points on an Argand plot of the modulation scheme, and $c_{max}$ is the largest possible real part of a value in the modulation scheme.

To recover the symbols after transmission, we define $\Theta_{\tau} (x) = \tau \lfloor x/ \tau + \frac{1}{2} \rfloor$ and $\Gamma_{\tau}(x) = x - \Theta_{\tau} (x)$ to represent the integer and fractional parts of the $\mathbf{u}$ vector with respect to $\tau$. The functions act element-wise on a vector. The recovered signal of
\begin{equation}
        \bar{\mathbf{u}} = \mathbf{u} + \Gamma_{\tau}\left(\sqrt{\frac{\lVert \mathbf{x} \rVert^2}{\rho}} \mathbf{n} \right)
\end{equation}
demonstrates the reduced impact of noise, when $\lVert \mathbf{x} \rVert^{2}$ is minimized. This yields an instance of the closest vector problem
\begin{equation}
        \label{eq_CVP}
        \mathbf{l}_{\mathrm{opt}} = \underset{\mathbf{l}}{\operatorname{argmin}} \lVert \mathbf{G}(\mathbf{u}- \tau \mathbf{l}) \rVert^{2},
\end{equation}
which is well known to be an NP hard problem.

\vspace{-0.3cm}
\section{Quantum Annealing}

The QA algorithm is an instance of adiabatic quantum computing that solves a specific set of problems, namely quadratic unconstrained binary optimization (QUBO) ~\cite{kadowaki_quantum_1998}. Inspired by the adiabatic theorem of quantum mechanics, QA finds the lowest energy configuration of a programmable magnetic quantum spin network that can be described by the Ising model. In QA, an appropriately selected initial Hamiltonian $\mathcal{H}_I$ is slowly evolved into the problem Hamiltonian $\mathcal{H}_P$ that maps the QUBO onto the spin states $\lvert -1\rangle$ and $\lvert +1 \rangle$.

The system commences its evolution from the ground state of
\begin{equation}
        \mathcal{H}_I = \sum_{i \in \mathsf{V}} \sigma_x^i,
\end{equation}
where $\sigma_x^i$ is the Pauli spin $x$ operator acting on the $i$th qubit; $\mathsf{V}$ is the set of qubits. This represents the application of a transverse magnetic field. The final measurements of spin are then taken along the $z$ axis, thus the qubits start in an equi-probable superposition of the spin states. The problem is encoded into the magnetic biases $h_{i}$ and couplings between the qubits $J_{ij}$ along the $z$ axis. This yields a problem Hamiltonian of the form
\begin{equation}
        \mathcal{H}_P = \sum_{i \in \mathsf{V}} h_i \sigma_z^i + \sum_{\langle i, j \rangle \in \mathsf{E}} J_{ij} \sigma_z^i \sigma_z^j,
\end{equation}
where $\mathsf{E}$ is the set of physical couplings between qubits $i$ and $j$. The adiabatic theorem states that a quantum system will remain in its ground state, provided that it transitions to a different Hamiltonian slowly enough. The full time-dependent Hamiltonian is formulated as
\begin{equation}
        \label{eq_Hamiltonian}
        \mathcal{H} = A(s) \mathcal{H}_I + B(s) \mathcal{H}_P,
\end{equation}
where $s = t/T_a$ and the function $A(s)$ and $B(s)$ are subject to the conditions that $A(0) = B(1) = 1$ and $A(1) = B(0) = 0$, while $T_a$ is the annealing time which must be long enough for the adiabatic theorem to hold. If $T_a$ is too short, the probability of remaining in the ground state is reduced, making the computation less likely to give an optimal solution ~\cite{kadowaki_quantum_1998}.

The spin variables may be transformed onto binary variables as follows $s_i = 2q_i -1$. Our QUBO may be represented as
\begin{equation}
        \mathbf{q}_{opt} =  \underset{\mathbf{q \in \mathbb{B}^{M}}}{\operatorname{argmin}} \quad \mathbf{q}^\top\mathbf{Q}\mathbf{q},
\end{equation}
where the legitimate values of the scalar $\mathbf{q}^\top\mathbf{Q}\mathbf{q}$ are related to the energies of the quantum spin network, while $\mathbb{B} = \{0, 1\}$, and $\mathbf{q}$ has a length of $M$. Since we have $q_i^2 = q_i$, the magnetic bias terms ($h_i$) are absorbed into the diagonal of $\mathbf{Q}$.

This work focuses on integer programming where the space of legitimate solutions is infinite i.e. any integer vector. Legitimate solutions of a QUBO take the form of any binary vector of a fixed dimension, this space is unconstrained but finite. Thus, the integer solution space is itself constrained to a finite number of vectors. This leads to the use of an encoding where the integer vector
\begin{equation}
        \label{eq_encoding}
        \mathbf{l} = \mathbf{C}\mathbf{q}.
\end{equation}
The encoding matrix $\mathbf{C} \in \mathbb{Z}^{N \times M}$ is a block matrix containing integers, where each element in $\mathbf{q} \in \mathbb{B}^{M}$ represents whether the integer in $\mathbf{C}$ is part of the sum that yields an element in $\mathbf{l} \in \mathbb{Z}^N$. Each column of $\mathbf{C}$ contains only a single integer. Encodings may be degenerate, i.e. an integer may not have a unique representation in binary form. The more integers that are encoded, the higher the qubit requirement becomes, and the higher the number of legitimate energy levels of the Ising model Hamiltonian becomes. The energy range of the QA hardware is fixed, and the problem values are scaled to fit into this range. Hence, for a higher number of energy levels, there is also a lower expected energy gap between the ground state and the excited states of the system, making the annealing more susceptible to noise as a consequence of the adiabatic theorem ~\cite{fetter_theoretical_2003}. To compensate for the reduced energy level gaps, $T_a$ must be increased to ensure a higher probability of measuring an optimal solution, however this comes at the cost of delicate quantum state being perturbed by environmental noise. This decoherence of the quantum state ~\cite{denchev_what_2016} and qubit manufacturing imperfections ~\cite{rajak_quantum_2023} are inherent sources of error to QA.

Once the QUBO has been formulated, it may be represented by a graph structure of logical variables. The physical hardware however already has a fixed structure, so the problem graph must be embedded in the hardware graph as a graph minor, where chains of physical qubits represent a single logical qubit. The problem is then transformed to its Ising formulation $\mathcal{H}_P$, and $\mathcal{H}$ evolves according to (\ref{eq_Hamiltonian}). The chains are positively coupled ($J_{ij} \ge 0$), ensuring that neighbouring spins remain aligned.

\vspace{-0.4cm}
\section{Algorithm Design}

The problem formulated in (\ref{eq_CVP}) is calculated in terms of complex variables. To encode the integers in the problem we rewrite the variables using the isomorphism between $\mathbb{C}$ and $\mathbb{R}^2$,
\begin{equation}
        \mathbf{M} \rightarrow \begin{pmatrix}
                \mathfrak{Re}[\mathbf{M}] & -\mathfrak{Im}[\mathbf{M}] \\
                \mathfrak{Im}[\mathbf{M}] & \mathfrak{Re}[\mathbf{M}] 
        \end{pmatrix}
         \qquad
        \mathbf{v} \rightarrow \begin{pmatrix}
                \mathfrak{Re}[\mathbf{v}] \\ \mathfrak{Im}[\mathbf{v}]
        \end{pmatrix},
\end{equation}
where $\mathbf{M}$ and $\mathbf{v}$ are an arbitrary complex matrix and vector. This is used to rewrite (\ref{eq_CVP}) in terms of only real variables, allowing for the encoding of real integers, and the lattice reduction algorithm to act as expected over a real valued lattice basis.

\begin{figure}[!t]
\centering
\includegraphics[width=0.85\columnwidth, trim={0 1.5cm 0 0}, clip]{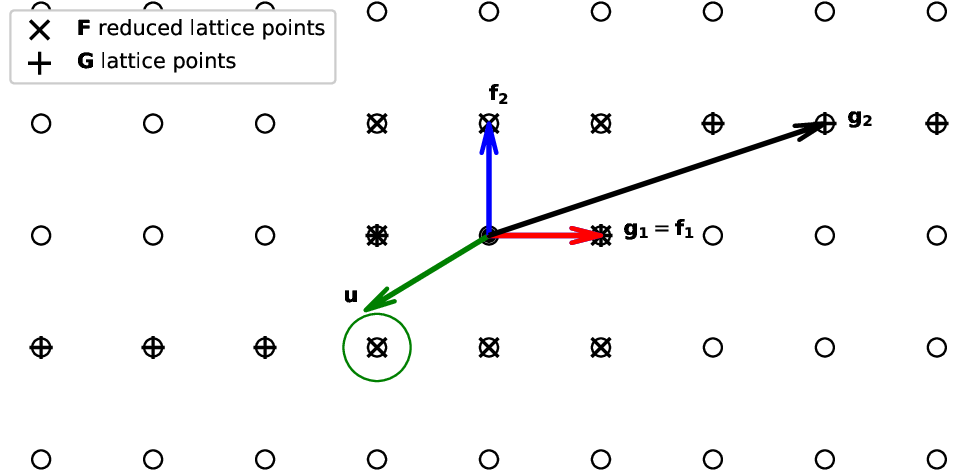}
\caption{The impact of lattice reduction on limited search spaces in 2D. The bases $\{\mathbf{g}_1, \mathbf{g}_2\}$ and $\{\mathbf{f}_1, \mathbf{f}_2\}$ describe the same lattice. The $\mathbf{u}$ vector in green is closest to the encircled lattice point. The search points in the non-reduced basis do not include this closest point, showing the necessity of lattice reduction in limited search space examples.}
\label{LatticePoints}
\end{figure}

Due to the limited resource of NISQ devices, the problem formulation and preprocessing are vital for generating useful results. Again, the LLL algorithm ~\cite{lenstra_factoring_1982} is employed for generating a short basis on which to solve the CVP. The matrix $\mathbf{G}$ can be said to form a lattice basis having columns that are the basis vectors. In the context of Figure \ref{LatticePoints}, the original basis is $\{\mathbf{g}_1, \mathbf{g}_2\}$. The LLL algorithm factors the original lattice $\mathbf{G} = \mathbf{F}\mathbf{Z}^{+}$ into the reduced lattice $\mathbf{F}$ and the unimodular matrix $\mathbf{Z}$. Another property of $\mathbf{F}$ is that of having a reduced Frobenius norm $\lVert \mathbf{F} \rVert_{F} \leq \lVert \mathbf{G} \rVert_{F}$. LLL reduces the lengths of the basis vectors, ensuring short vectors are typically represented by small integers. In the context of Figure \ref{LatticePoints}, $\{\mathbf{f}_1, \mathbf{f}_2\}$ is the reduced basis. To formulate the CVP in the reduced basis, let 
\begin{align}
        \mathbf{u'} &= \Gamma_\tau(\mathbf{Z}^+ \mathbf{u}) \\
        \mathbf{l'} &= \mathbf{Z}^+ + \mathbf{l} \Theta_\tau(\mathbf{Z}^+ \mathbf{u})
\end{align}
and then substitute them into (\ref{eq_x_def}), which yields
\begin{equation}
        \label{eq_reduced_x}
        \mathbf{x} = \mathbf{F}(\mathbf{u'} - \tau \mathbf{l'}),
\end{equation}
noting that in the case of $\mathbf{l}=\mathbf{0}$ it is equivalent to the LRZFP.

The elements of the integer perturbation vector are typically in the set $\{-1, 0, 1\}$ ~\cite{yuen_wlcp1-13_2006}. This, along with the lattice reduction, justifies the choice to only encode the integers in this set. To encode the integers, the scheme laid out in ~\cite{karimi_practical_2019} is used, but this is only valid for positive integers. A translation is used for rectifying the mismatch between the two sets allowing the search space to shift to $\{0, 1, 2\}$, where $\mathbf{u''} = \mathbf{u'} + \tau \mathbf{t}$ and $\mathbf{l''} = \mathbf{l'} + \mathbf{t}$ with $\mathbf{t} = \begin{pmatrix}
        1 & 1 & \hdots & 1
\end{pmatrix}^\top$. The problem can then be transformed into a QUBO accordingly. We encode a single integer,
\vspace{-0.2cm}
\begin{equation}
        l_i = \begin{pmatrix}
                1 & 1
        \end{pmatrix}
        \begin{pmatrix}
                q_{i0} \\
                q_{i1}
        \end{pmatrix},
\vspace{-0.16cm}
\end{equation}
showing that the integer `0' is represented in binary form as `00`', while `1' has two representations - `01' and `10' - and `2' is represented as `11'. The redundancy in encoding reduces the number of energy levels in $\mathcal{H}$ by introducing degeneracy, making the encoding more resilient to noise. The smaller number of energy levels will be scaled to the annealer's hardware capabilities, leaving a larger energy gap between the ground and excited states of $\mathcal{H}$ compared to that of a more dense encoding, hence allowing for reduced annealing times.

To transform the problem into a QUBO, we substitute (\ref{eq_encoding}) and (\ref{eq_reduced_x}) into (\ref{eq_CVP}), followed by expanding it, yielding
\begin{align}
        \begin{split}
                \label{eq_qubo_def}
        \lVert \mathbf{x} \rVert^{2} ={}& \tau^{2} \mathbf{q}^\top \mathbf{C}^\top\mathbf{F}^\top\mathbf{F}\mathbf{C}\mathbf{q} - 2 \tau \mathbf{u''}^\top \mathbf{F}\mathbf{C}\mathbf{q} \\
        &+ \mathbf{u''}^\top\mathbf{F}^\top\mathbf{F}\mathbf{u''} 
        \end{split}\\
        ={}& \mathbf{q}^\top \mathbf{Q} \mathbf{q} + c,
\end{align}
where we have $\mathbf{Q} = \tau^{2} \mathbf{C}^\top \mathbf{F}^\top \mathbf{F}\mathbf{C} + \operatorname{diag}(- 2 \tau \mathbf{u''}^\top \mathbf{F}\mathbf{C})$ and $c = \mathbf{u''}^\top\mathbf{F}^\top\mathbf{F}\mathbf{u''}$. The $\operatorname{diag}(\cdot)$ function takes the row vector argument and places its elements on the diagonal of a matrix. 

The graph $\mathbf{Q}$ represents is typically complete for this problem. The physical connections of qubits on the quantum annealer do not match up with the structure, thus the QUBO must be embedded in the hardware graph, where chains of physical qubits represent a single logical qubit. The `minor miner' algorithm ~\cite{cai_practical_2014} generates a new graph that matches the hardware connections, and an implementation can be found in the Ocean SDK ~\cite{OceanSDK}. The chains require a positive coupling ($J_{chain} > 0$) so that the spins remain aligned. To assign the chain strength, the uniform toque compensation function in the Ocean SDK is used that calculates,
\begin{equation}
        \label{eq_chain_strength}
        J_{chain} = p \cdot \sqrt{\sum_{\langle i, j \rangle \in \mathsf{E}} \frac{J_{ij} ^{2}}{N_J}} \cdot \sqrt{k_{avg}},
\end{equation}
where $k_{avg}$ is the average degree of a node in the problem graph, and $N_J$ is the number of off diagonal elements of $\mathbf{Q}$. The chain strength parameter $p$ may be varied to find the optimal performance of the algorithm. At this stage, $\mathcal{H}_P$ has been formulated and may be solved with the QA algorithm. The algorithm framework is summarized in Alg. \ref{QA_VP}
\begin{algorithm}
\footnotesize
\caption{QA-based LR-VP Precoding}  
\label{QA_VP}
\Input{$\mathbf{H}$, $\mathbf{u}$} \Output{$\mathbf{l''}$}
Invert $\mathbf{H}$ and apply the LLL algorithm ~\cite{lenstra_factoring_1982} to generate $\mathbf{F}$ and $\mathbf{Z}$.\\
Generate the encoding matrix $\mathbf{C}$ and calculate the QUBO using (\ref{eq_qubo_def}).\\
Embed the QUBO in the hardware graph and calculate $J_{chain}$ according to (\ref{eq_chain_strength}).\\
Apply the QA algorithm and invert the binary result back into an integer vector according to (\ref{eq_encoding}).\\
\end{algorithm}

%

\vspace{-0.38cm}
\section{Results}

The experimental parameters used in the section are summarized in Table \ref{Parameters}, unless explicitly stated that they different. The number of transmitted symbols applies only to the symbol error rate (SER) simulations.
\begin{table}[!ht]
    \centering
    \caption{Experimental parameters}
    \label{Parameters}
    \begin{tabular}{|l|l|}
    \hline
        Parameter & Value \\ \hline
        Encoding & 64 QAM \\ \hline
        $N_r \times N_t$ & 10$\times$10 \\ \hline
        Number of transmitted symbols & 150 000 \\ \hline
        $T_a$ & 100 $\mu$s \\ \hline
        $p$  & 1.5 \\ \hline
    \end{tabular}
\end{table}

\begin{figure}[!t]
\centering
\includegraphics[width=0.8\columnwidth, trim={0 0.6cm 0 1.0cm},clip]{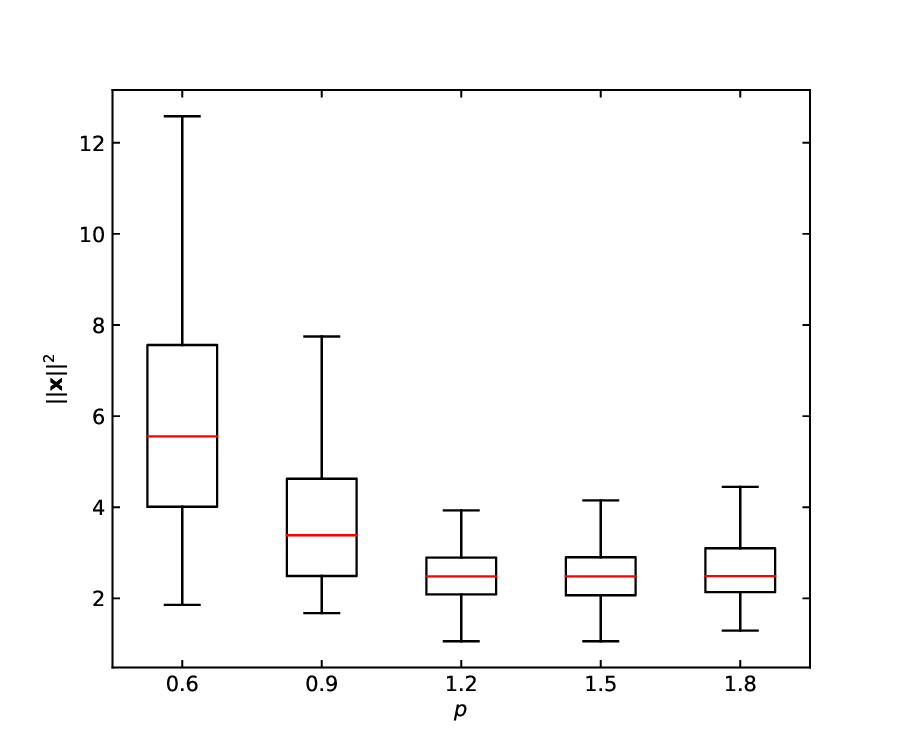}
\caption{Distributions of $\lVert \mathbf{x} \rVert^2$, calculated via the LRAQVP algorithm, with varying chain strength parameter. The reduced median values show the impact of increased chain strength counteracting the noise, ensuring better solution quality. The limited dynamic range of the quantum annealer becomes the dominant factor in noise above the optimal value of $p$. There are 150 samples in each distribution.}
\label{fig_chain_strength}
\end{figure}

Tuning $T_a$ and $p$ are important steps in order to ensure optimal performance of QA. Furthermore, the dependence on these parameters gives an insight into the robustness of the algorithm against noise and the limited dynamic range of the hardware. In total 150 channels and signals were generated at random and (\ref{eq_CVP}) was solved using the DWave Advantage quantum annealer, with $T_a$ varying from 20 $\mu$s to 400 $\mu$s and $p$ fixed at 1.5. We observed that, without showing graphically due to page limit, the choice of $T_a$ had no impact on the distributions of $\lVert \mathbf{x} \rVert^2$, indicating that there is a robust production of a large enough energy gap between the ground and first excited states of $\mathcal{H}$. The lack of an increase in median values with increasing $T_a$ shows that other sources of noise are not impacting the quality of the results. For the rest of the experiments $T_a$ is fixed at 100 $\mu$s as a conservative value, selected for reducing the likelihood of measuring suboptimal solutions. The optimization of $p$ was carried out similarly with $p$ varying between 0.6 and 1.8. The distributions of $\lVert \mathbf{x} \rVert^2$ are shown in Fig. \ref{fig_chain_strength}, where the optimal median value lies at $p=1.5$. The large median and variance at low values of $p$ are due to broken chains leading to suboptimal results being calculated however, when $p \ge 1.2$ we see a compression in the distributions indicating robust chains and reliable solutions. The increase of the median value at $p=1.8$ is due to the limited dynamic range of the qubit couplers. All values in $\mathcal{H}_P$ get scaled to fit the range of the hardware, thus the larger value of $J_{chain}$ lead to a compression of gaps between the energy levels of the problem leading to a greater susceptible to noise. The distributions of $\lVert \mathbf{x} \rVert^2$ shown in Fig. \ref{fig_chain_strength} are displayed with the outliers removed. The outliers are those that lie outside 1.5 times the interquartile range (the height of the boxes in Fig. \ref{fig_chain_strength}), above the upper quartile, and below the lower quartile.

\begin{figure}[!t]
\centering
\includegraphics[width=0.8\columnwidth, trim={0 0.6cm 0 1.0cm},clip]{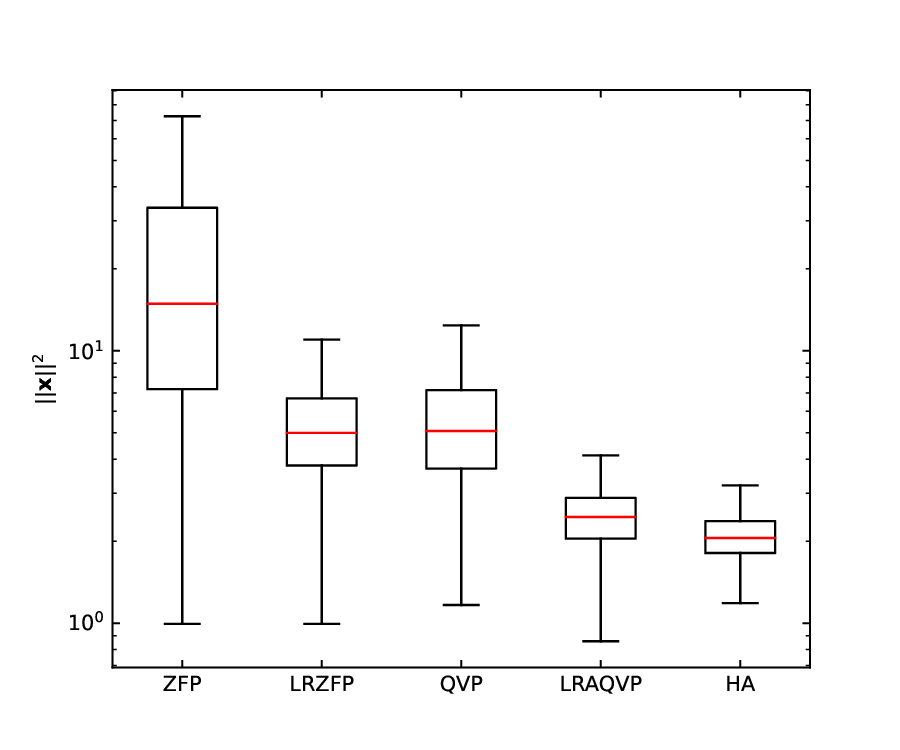}
\caption{The distributions of $\lVert \mathbf{x} \rVert^{2}$ for Zero forcing, the LRZFP protocol and the LRAQVP presented here. The HA lower bound represent a spherical approximation to what the closest vector could be. The HA is a lower bound on the median performance that is averaged over all symbol vectors that could pass through a given channel.}
\label{Distributions}
\end{figure}

Fig. \ref{Distributions} shows performance in terms of the optimization objective, several algorithms were compared over a sample size of 2000 random channels and signals.  The Hyper-sphere Approximation (HA) is calculated for each channel according to Lemma 1 in ~\cite{ryan_performance_2009}. The HA represents a lower bound of the SER for the average of all the symbols under a sphere encoder of exponentially escalating complexity. QVP was found to have a similar distribution to LRZFP, but LRAQVP is approaching the HA in terms of its median performance. In all the channels, a non-zero perturbation vector $\mathbf{l}$ was found by the LRAQVP algorithm.

\begin{figure}[!t]
\centering
\includegraphics[width=0.8\columnwidth, trim={0 0.3cm 0 0.7cm},clip]{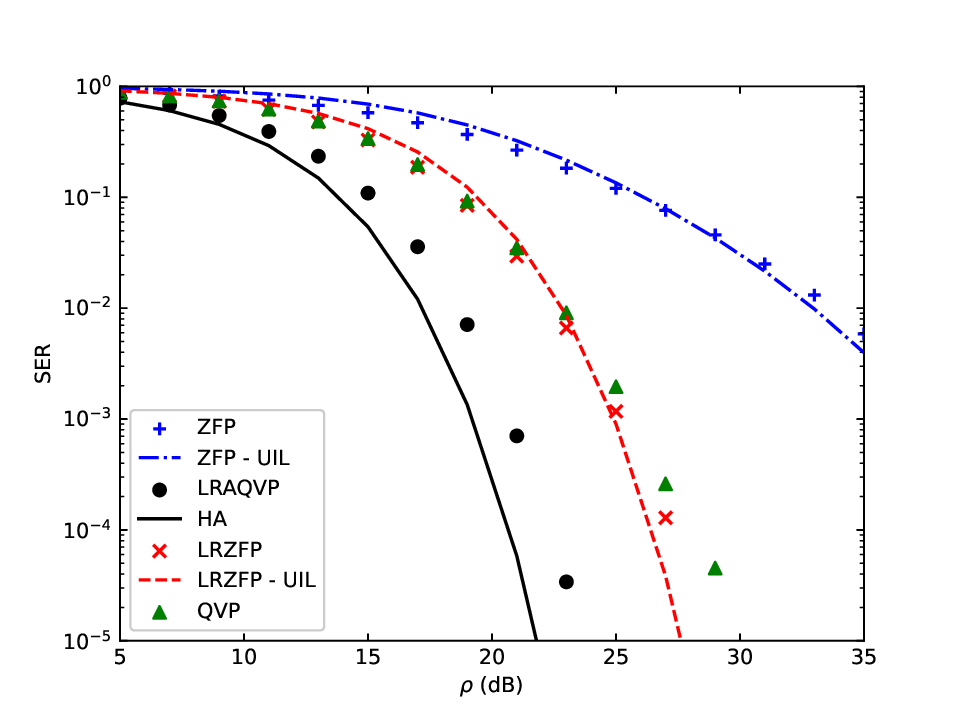}
\caption{The SER simulations for a 10x10 MIMO system. A 5 dB gain can be seen from LRZFP to LRAQVP protocols. QVP can be seen to behave similarly to LRZFP, however it scales with much higher computational complexity, and performs worse in the high SNR regime. The uniform input limits (UILs) are theoretical predictions for the zero forcing protocols used to verify the empirical results. HA is a lower bound on the performance, that LRAQVP is approaching.}
\label{SER}
\end{figure}

SER simulations were carried out to compare the algorithms in a communications system. We also compare the empirical results to the theoretical uniform input limit (UIL). The prediction for LRZFP has the associated shaping loss compensated for ~\cite{honig_capacityapproaching_2009}. Despite being valid only for large $\rho$, it fits the empirical data well. $T_a$ and $p$, were held constant at 100 $\mu$s and 1.5 for LRAQVP and QVP. Since LRAQVP is expected to be a more reliable algorithm, the annealing time may be longer than it is strictly necessary. Lastly, the HA prediction constitutes a lower bound on the expected SER of the sphere encoder ~\cite{ryan_performance_2009}. For LRAQVP and QVP, each perturbation vector was calculated using the DWave Advantage. The end to end SER performance is much improved over that of LRZFP, and approaches the HA. There is an approximately 5 dB gain over LRZFP. The predictions and the results of the SER simulations are shown in Fig. \ref{SER}. 


The complexities of the core algorithms discussed are displayed in Table \ref{ComplexityTable}. Explicitly, the complexity of QA is governed by the annealing time required for the adiabatic theorem to hold ~\cite{mukherjee_multivariable_2015}. During the annealing, there is only one evaluation of $\lVert \mathbf{x} \rVert^2$, whilst for a classical algorithm, $\lVert \mathbf{x} \rVert^2$ is iteratively evaluated. The number of evaluations dictates its complexity. The upper bound on the number of evaluations of $\lVert \mathbf{x} \rVert^2$, for a sphere encoder, is an exponential function of the number of users \mbox{~\cite{ryan_performance_2009}}. The complexity analysis shows that there exists some $N_r$ at which it would be faster to use a quantum annealer to find a solution to the CVP compared to a sphere encoder. The total computation time for the QA section of LRAQVP, for a single channel and signal vector, was 260 ms as 1000 anneals were performed to maximize the likelihood of measuring the ground state. The 1000$\times$100 $\mu$s $=$ 100 ms of quantum annealing accounts for solving the problem and the remainder is the time required for the chip to reset and for the temperature to reduce again. The typical coherence times for 5G radio channels are on the order of milliseconds ~\cite{zhang_cell_2023}, but they can be shorter depending on the mobility of a user. Therefore, the impact of noise within QA must be much reduced so that the number of anneals required is reduced to reach a practical timescale. A solution may be to pair QA with channel stabilization techniques, increasing the channel's coherence time. For much larger antenna arrays and cell-free communications, QA may be viewed as a door into nonlinear TPC on a massive scale.

\vspace{-0.3cm}
\begin{table}[!ht]
    \centering
    \caption{Time complexity in number of users.}
    \label{ComplexityTable}
    \begin{tabular}{|l|l|}
    \hline
        Algorithm & Complexity \\ \hline
        ZFP & $\mathcal{O}(N_r^{2.373})$ \\ \hline
        LRZFP & $\mathcal{O}(N_r^{5})$ \\ \hline
        Sphere Encoder & $\mathcal{O}(e^{N_r})$ \\ \hline
        Quantum Annealing & $\mathcal{O}(e^{\sqrt{N_r}})$ \\ \hline
    \end{tabular}
\end{table}

\vspace{-0.8cm}
\section{Conclusions}

The problem formulation and preprocessing in this LRAQVP framework has been shown to give good performance for a quantum CVP solver in MIMO systems. Defining the problem in terms of a short lattice basis allows searching over a limited subset of the most relevant points on the lattice, with small integers available for encoding on limited quantum hardware and paving the way for more practical implementations relying on NISQ annealing hardware. The performance of the LRAQVP framework has been experimentally validated on real QA hardware, and it was also benchmarked against the classical LRZFP protocol. As all the perturbation vectors were non-zero, this underlines the importance of an effective CVP solver, even in conjugation with lattice reduction. The complexity of classical algorithms demonstrates that QA-based protocols could outperform them for larger problem sizes, while still being practically feasible. However, optimal preprocessing of the problem demonstrates the importance of lattice reduction in ensuring quantum advantage over classical algorithms.

\vspace{-0.7cm}
\bibliography{refs}
\bibliographystyle{IEEEtran}

\newpage

\vfill

\end{document}